\documentclass[11pt]{article}

\usepackage{color,amsmath,amssymb,epsf,epsfig,subfigure,psfrag,afterpage,chicago}

\newcommand{\bfzero   }{\hbox{\boldmath$0$}}

\newcommand{\bftheta  }{\hbox{\boldmath$\theta$}}

\newcommand{\bfepsilon}{\hbox{\boldmath$\epsilon$}}

\newcommand{\bfgamma  }{\hbox{\boldmath$\gamma$}}

\newcommand{\bfD}{\hbox{\boldmath$D$}}

\newcommand{\bfH}{\hbox{\boldmath$H$}}
\newcommand{\bfI}{\hbox{\boldmath$I$}}

\newcommand{\bfQ}{\hbox{\boldmath$Q$}}

\newcommand{\bfR}{\hbox{\boldmath$R$}}

\newcommand{\bfu}{\hbox{\boldmath$u$}}

\newcommand{\bfx}{\hbox{\boldmath$x$}}

\newcommand{\bfy}{\hbox{\boldmath$y$}}

\newcommand{\s    }{\sigma^2}

\begin{document}

\author{Jos\'e A. Fioruci, Ricardo S. Ehlers, Francisco Louzada}

\title{BayesDccGarch - An Implementation of Multivariate GARCH DCC Models}

\maketitle

\begin{abstract}
  
Multivariate GARCH models are important tools to describe the dynamics
of multivariate times series of financial returns.
Nevertheless, these models have been much less used in practice due to
the lack of reliable software.
This paper describes the {\tt R} package {\bf BayesDccGarch} which was
developed to implement recently proposed inference procedures to
estimate and compare multivariate GARCH models allowing for asymmetric
and heavy tailed distributions.
\vskip .5cm


\end{abstract}

\section{Introduction}

Modelling the dynamics of financial returns has been object of much
attention and extensively researched for decades since the
introduction of the Autoregressive Conditional Heteroskedasticity
(ARCH) model in the seminal work of \citeN{engle82} and its
generalization, the GARCH model of \citeN{boll86}. However, when
dealing with multivariate returns, one must take into account the
mutual dependence between them. For example, the dependence between
assets tends to increase in periods of market turbulence which in turn
might have implications for portfolio and risk management. Analyzing
asset return covariances is crucial to portfolio selection, asset
management and risk assessment.

It is therefore natural to extend to multivariate GARCH (MGARCH)
models and this is the main focus of this paper. For recent reviews on
multivariate GARCH models see for example \shortciteN{baulr06},
\citeN{sil-tera09} and \citeN{tsay10}. It is important that the model
is flexible enough to be able to capture the dynamics in the
conditional variances and covariances and yet parsimonious enough so
that parameter estimation does not come at a high computational cost.
In this paper, we focus on the so called
conditional correlation models which allow to specify separately the
individual conditional variances and the conditional correlation
matrix. In this specification it is straightforward to impose positive
definiteness in the conditional variance covariance matrix using some
simple constraints.

The fact that the distribution of financial returns are typically
characterized by fat tails and frequently show some degree of
asymmetry should be taken into consideration too. To introduce some
degree of asymmetry, \shortciteN{cappuccio04} adopted a Bayesian
approach to estimate stochastic volatility models with a skewed power
exponential distribution while \citeN{pipien06} and \citeN{ehl2012}
compared GARCH models with different skewed distributions for the
error terms.
\citeN{baul05} proposed a
new class of skew multivariate Student $t$ distributions and parameter
estimation was based on (quasi-) maximum likelihood methods.
\shortciteN{ehl2014} adopted a similar modelling approach from a
Bayesian perspective and included a skew multivariate GED distribution.
\shortciteN{ausin14a} proposed a Bayesian non-parametric approach with
applications to portfolio selection.

We give an overview of multivariate GARCH models in Section
\ref{sec:mgarch} using both symmetric and skewed multivariate
distributions. An overview of the package implementation and usage
including illustration with real data is provided in Section
\ref{sec:pac}. Finally Section \ref{sec:conclusion} offers some
conclusions and possible future steps in improving the implementation.

\section{A Multivariate GARCH Model}\label{sec:mgarch}

Consider a multivariate time series of returns
$\bfy_t=(y_{1t},\dots,y_{kt})'$ such that $E(\bfy_t)=\bfzero$. The
model assumes that $\bfy_t$ is conditionally heteroskedastic,
\begin{equation}\label{eq:eq1}
\bfy_t = \bfH_t^{1/2}\bfepsilon_t  
\end{equation}
where $\bfH_t^{1/2}$ is any $k\times k$ positive definite matrix such
that the conditional variance of $\bfy_t$ is $\bfH_t$ and which depends on a
finite vector of parameters $\bftheta$. The $k\times 1$ error vectors
are assumed independent and identically distributed with
$E(\bfepsilon_t)=0$ and $E(\bfepsilon_t\bfepsilon_t')=\bfI_k$, where $\bfI_k$ is the
identity matrix of order $k$. 

There are different possible specifications for $\bfH_t$. In this
paper, we focus on the so called 
conditional correlation models which allow to specify separately the
individual conditional variances and the conditional correlation
matrix. \citeN{boll90} proposed a parsimonious approach in
which the conditional covariances are proportional to the product of
the corresponding conditional standard deviations. The constant
conditional correlation model is defined as,
$\bfH_t=\bfD_t \bfR \bfD_t$, 
where $\bfD_t=\mbox{diag}(h_{11,t}^{1/2},\dots,h_{kk,t}^{1/2})$, $\bfR$ is a
symmetric positive definite matrix which elements are the conditional
correlations $\rho_{ij}$, $i,j=1,\dots,k$. Each conditional covariance
is then given by $h_{ij,t}=\rho_{ij} \sqrt{h_{ii,t}~h_{jj,t}}$.
Besides, each conditional variance in $\bfD_t$ is specified as a univariate GARCH model.
Here we specify a GARCH(1,1) model for each conditional variance, i.e.
\begin{equation}\label{eq:garch11}
h_{ii,t}=\omega_i +\alpha_i y_{i,t-1}^2 + \beta_i h_{ii,t-1}, ~i=1,\dots,k,
\end{equation}
with $\omega_i>0$, $\alpha_i\ge 0$,
$\beta_i\ge 0$ and
$\alpha_i+\beta_i < 1$, $i=1,\dots,k$.
Note that this model contains $k(k+5)/2$ parameters. It is clear that
$\bfH_t$ is positive definite if and only if $h_{ii,t}>0, i=1,\dots,k$
and $\bfR$ is positive definite.

\citeN{engle02}, \citeN{chris02} and \citeN{tse02} independently
proposed generalizations by allowing the
conditional correlation matrix to be time dependent. The resulting
model is then called a dynamic conditional correlation (DCC) MGARCH
model. We adopted the same approach in
\citeN{engle02} by setting the following parsimonious formulation for
the correlation matrix,
\begin{equation}\label{eq:rt}
\bfR_t=\mbox{diag}(\bfQ_t)^{-1/2} ~\bfQ_t ~\mbox{diag}(\bfQ_t)^{-1/2},
\end{equation}
where $\bfQ_t$ are $k\times k$ symmetric positive-definite matrices given
by,
\begin{equation}\label{eq:qt}
\bfQ_t=(1-\alpha-\beta)\bfR +\alpha \bfu_{t-1} \bfu_{t-1}' +\beta \bfQ_{t-1},  
\end{equation}
$\bfu_t=\bfD_t^{-1}\bfy_t$ are the standardized returns and $\bfR$ is the
unconditional covariance matrix of $\bfu_t$. Also,
$\alpha>0$, $\beta>0$ and $\alpha+\beta<1$. After some algebra it is
not difficult to see that the conditional covariances are given by
$ h_{ij,t}=q_{ij,t} \sqrt{h_{ii,t}h_{jj,t}} /
\sqrt{q_{ii,t}q_{jj,t}}$. So, from (\ref{eq:rt}) and (\ref{eq:qt}) we
can see that $\bfQ_t$ is written as a GARCH(1,1)-type equation and
then transformed to give the correlation matrix $\bfR_t$. 

\subsection{Skewed Distributions}\label{sec:skew}

\citeN{baul05} proposed to construct a multivariate skew distribution
from a symmetric one by changing the scale on each side of the mode
for each coordinate of the multivariate density. This is a multivariate
extension of what \citeN{fsteel98} had proposed as a skewing mechanism
for univariate distributions. The density is given by,

\begin{equation}\label{eq:mskew}
s(\bfx\vert\bfgamma)=
2^k\left(\prod_{i=1}^{k}\frac{\gamma_i}{1+\gamma_i^2}\right)f(\bfx^*)
\end{equation}
where $f(\cdot)$ is a symmetric multivariate density,
$\bfx^*=(x_1^*,\dots,x_k^*)$ such that $x_i^*=x_i/\gamma_i$ if $x_i\ge
0$ and $x_i^*=x_i\gamma_i$ if $x_i< 0$, $i=1,\dots,k$. The parameters
$\gamma_i$ control the degree of skewness on each margin, right (left)
marginal skewness corresponding to $\gamma_i>1$ ($\gamma_i<1$). Also,
the existence of the moments of Equation (\ref{eq:mskew}) depends only
on the existence of the marginal moments $E(X_i^r)$ in the original
symmetric distribution. The interpretation of each $\gamma_i$ is the
same as in \citeN{fsteel98}, i.e. $\gamma_i^2 = Pr(X_i\ge 0)/Pr(X_i<0)$
which helps with the specification of a prior distribution. Besides
the multivariate skew normal distribution we allow the error term to follow
a multivariate skew $t$ or a multivariate skew GED as well. In all
cases, setting $\gamma_i=1$, $i=1,\dots,k$ recovers the original
symmetric density.

\subsection{Prior Distributions}

The model specification is completed specifying prior distributions of
all parameters of interest. These are assumed to be a priori
independent and normally distributed truncated to the intervals that
define each one (\shortciteNP{ehl2014}). 
For the GARCH(1,1) coefficients in (\ref{eq:garch11}), these prior
distributions are the same as proposed in \citeN{ardia06}, 
$\omega_i\sim N(\mu_{\omega_i},\s_{\omega_i})I_{(\omega_i>0)}$, 
$\alpha_i\sim N(\mu_{\alpha_i},\s_{\alpha_i})I_{(0<\alpha_i<1)}$ and
$\beta_i\sim N(\mu_{\beta_i},\s_{\beta_i})I_{(0<\beta_i<1)}$, $i=1,\dots,k$.
The tail parameter is assigned a truncated normal distribution as
$\nu\sim N(\mu_{\nu},\s_{\nu})I_{(\nu>2)}$ or 
$\delta\sim N(\mu_{\delta},\s_{\delta})I_{(\delta>0)}$
when using the multivariate Student $t$ or GED respectively.
Finally, a similar approach is adopted for the parameters $\alpha$ and
$\beta$ in equation (\ref{eq:qt}), i.e. 
$\alpha\sim N(\mu_{\alpha},\s_{\alpha})I_{(0<\alpha<1)}$ and
$\beta\sim N(\mu_{\beta}, \s_{\beta})I_{(0<\beta<1)}$. 
As for the skewness parameters, we use truncated normal distributions
on positive values, i.e. 
$\gamma_i\sim N(\mu_{\gamma_i},\s_{\gamma_i})I_{(\gamma_i>0)}$, $i=1,\dots,k$.

\section{Package Implementation}\label{sec:pac}

The major function in package {\bf bayesDccGarch} is 
{\tt bayesDccGarch} for\linebreak Bayesian estimation of the DCC-GARCH(1,1)
model. The function performs simulations from the posterior
distribution via Metropolis-Hastings algorithm. The synopsis of the
function is, 

\begin{verbatim}
bayesDccGarch(mY, nSim = 10000, tail_ini = 8, omega_ini = rep(0.03, 
    ncol(mY)), alpha_ini = rep(0.03, ncol(mY)), beta_ini = rep(0.8, 
    ncol(mY)), a_ini = 0.03, b_ini = 0.8, gamma_ini = rep(1,
    ncol(mY)), errorDist = 2, control = list())
\end{verbatim}
where the arguments are,

\begin{itemize}
\item {\tt mY}: a matrix of the data ($n\times k$).
\item {\tt nSim}: length of Markov chain. Default: 10000.
\item {\tt tail\_ini}: initial value of parameter $\nu$ if {\tt errorDist = 2} or
  initial value of parameter $\delta$ if {\tt errorDist = 3}. 
  If {\tt errorDist = 1} this arguments is not used. 
\item {\tt omega\_ini}: a numeric vector ($k\times 1$) with the initial values of
  $\omega_i$ parameters. Default: {\tt rep(0.03, ncol(mY))}.
\item {\tt alpha\_ini}: a numeric vector ($k\times 1$) with the initial values of
  $\alpha_i$ parameters. Default: {\tt rep(0.03, ncol(mY))}.
\item {\tt beta\_ini}: a numeric vector ($k\times 1$) with the initial values of
  $\beta_i$ parameters. Default: {\tt rep(0.8, ncol(mY))}.
\item {\tt a\_ini}: a numeric value of the initial values of parameter $a$.
  Default: 0.03.
\item {\tt b\_ini}: a numeric value of the initial values of parameter $b$.
  Default: 0.8.
\item {\tt gamma\_ini}: a numeric vector ($k\times 1$) with the initial values of
  $\gamma_i$ parameters. Default: {\tt rep(1.0, ncol(mY))}.
\item {\tt errorDist}: a probability distribution for errors. Use {\tt
  errorDist=1} for SSNorm, {\tt errorDist=2} for SST or {\tt errorDist=3} for SSGED.
  Default: 2.
\item {\tt control}: list of control arguments (See Details).
\item {\tt outList}: an output element of {\tt bayesDccGarch} function.
\end{itemize}

\noindent The {\tt control} argument can be used to specify the prior
hyper-parameters and the simulation algorithm parameters. This is a
list where the user can supply any of the following components,

\begin{itemize}
\item {\tt \$mu\_tail} the value of hyper-parameter $\mu_{\nu}$ if {\tt
  errorDist=2} or the hyper-parameter $\mu_{\delta}$ if {\tt errorDist=3}. Default: 8
\item {\tt \$mu\_gamma} a vector with the hyper-parameters $\mu_{\gamma_i}$.
  Default: {\tt rep(0,ncol(mY)}
\item {\tt \$mu\_omega} a vector with the hyper-parameters $\mu_{\omega_i}$.
  Default: {\tt rep(0,ncol(mY)}
\item {\tt \$mu\_alpha} a vector with the hyper-parameters $\mu_{\alpha_i}$.
  Default: {\tt rep(0,ncol(mY)}
\item {\tt \$mu\_beta} a vector with the hyper-parameters $\mu_{\beta_i}$. Default:
  {\tt rep(0,ncol(mY)}
\item {\tt \$mu\_a} the value of the hyper-parameter $\mu_a$. Default: 0
\item {\tt \$mu\_b} the value of the hyper-parameter $\mu_b$. Default: 0
\item {\tt \$sigma\_tail} the value of hyper-parameter $\sigma_{\nu}$ if {\tt errorDist=2}
  or the hyper-parameter $\sigma_{\delta}$ if {\tt errorDist=3}. Default: 10
\item {\tt \$sigma\_gamma} a vector with the hyper-parameters $\sigma_{\gamma_i}$.
  Default: {\tt rep(1.25,ncol(mY)}
\item {\tt \$sigma\_omega} a vector with the hyper-parameters $\sigma_{\omega_i}$.
  Default: {\tt rep(10,ncol(mY)}
\item {\tt \$sigma\_alpha} a vector with the hyper-parameters $\sigma_{\alpha_i}$.
  Default: {\tt rep(10,ncol(mY)}
\item {\tt \$sigma\_beta} a vector with the hyper-parameters $\sigma_{\beta_i}$.
  Default: {\tt rep(10,ncol(mY)}
\item {\tt \$sigma\_a} the value of the hyper-parameter $\sigma_a$. Default: 10
\item {\tt \$sigma\_b} the value of the hyper-parameter $\sigma_b$. Default: 10
\item {\tt \$simAlg} the random walk Metropolis-Hasting algorithm update. Use
  ``1'' for updating all parameters as one block, use ``2'' for
  updating one parameter at a time and use ``3'' for an automatic choice.
\item {\tt \$cholCov} the Cholesky decomposition matrix of the covariance
  matrix for simulation by one-block Metropolis-Hasting. It
  must be passed if {\tt control\$simAlg=1}.
\item {\tt \$sdSim} a vector with the standard deviations for simulation by
  one-dimensional Metropolis-Hasting. It must be passed if
  {\tt control\$simAlg=2}.
\item {\tt \$print} a logical variable for if the function should report the
  number of iterations every 100 iterations or not. Default: TRUE
\end{itemize}

\noindent After completing the simulations, {\tt bayesDccGarch()} returns an
object of class {\tt bayesDccGarch} which is a list containing the
following components, 
\begin{itemize}
\item {\tt \$control} : a list with the used {\tt control} argument.
\item {\tt \$MC} : an element of class {\tt mcmc} from package {\bf
  coda} with the Markov chain simulations of all parameters.
\item {\tt \$elapsedTime} : an object of class {\tt proc\_time} which is a
  numeric vector of length 5, containing the user, system, and total
  elapsed times of the process. 
\end{itemize}

We note that the package {\bf coda} is loaded automatically with a
call to {\bf bayesDccGarch}. This package contains functions
which provide tools for MCMC output analysis and diagnostics 
(\shortciteNP{plummer-etal06}).

\subsection{Run MCMC}

The sampler that is currently implemented in {\bf bayesDccGarch} is
a random-walk Metropolis algorithm in which the parameters are
sampled in one block from a multivariate normal proposal
distribution centered about the current value. By default, the
variance-covariance matrix is specified as the negative inverse
Hessian matrix evaluated at the posterior mode which is obtained
numerically. This is accomplished with a call to the {\tt R} function
{\tt optim()}. 

If the optimizer fails to find the mode or the Hessian
matrix at the mode is not positive definite a random-walk Metropolis
is run on each individual parameter sampling from a univariate normal
proposal centered about the current value with a variance that tunes
the acceptance rates to be between 0.20 and 0.50. A sample
variance-covariance matrix is then calculated from the output of this
algorithm which is in turn multiplied by a scale factor and taken as
the variance-covariance matrix of the multivariate normal proposal
distribution in the one-block random-walk Metropolis.
We next provide some illustrations on the use of the package.

\subsection{Illustrations}


In this section we illustrate the package usage with
the {\tt DaxCacNik} dataset included in the package. First load the
{\bf bayesDccGarch} package into {\tt R},

\begin{verbatim}
> library(bayesDccGarch)  
\end{verbatim}

\noindent and the dataset provided,

\begin{verbatim}
> data(DaxCacNik)  
\end{verbatim}

\noindent 
The {\tt DaxCacNik} dataset contains daily observations of the
log-returns of daily indices of stock markets in Frankfurt (DAX),
Paris (CAC40) and Tokyo (NIKKEI), from 10 October 1991 until 30
December 1997 (a total of 1627 observations). These data are also
freely available from the website 
http://robjhyndman.com/tsdldata/data/FVD1.dat.
\vskip .2cm

\noindent We begin by running a univariate analysis on the DAX index
with a skew Student $t$ distribution for the errors.

\begin{figure}
\includegraphics{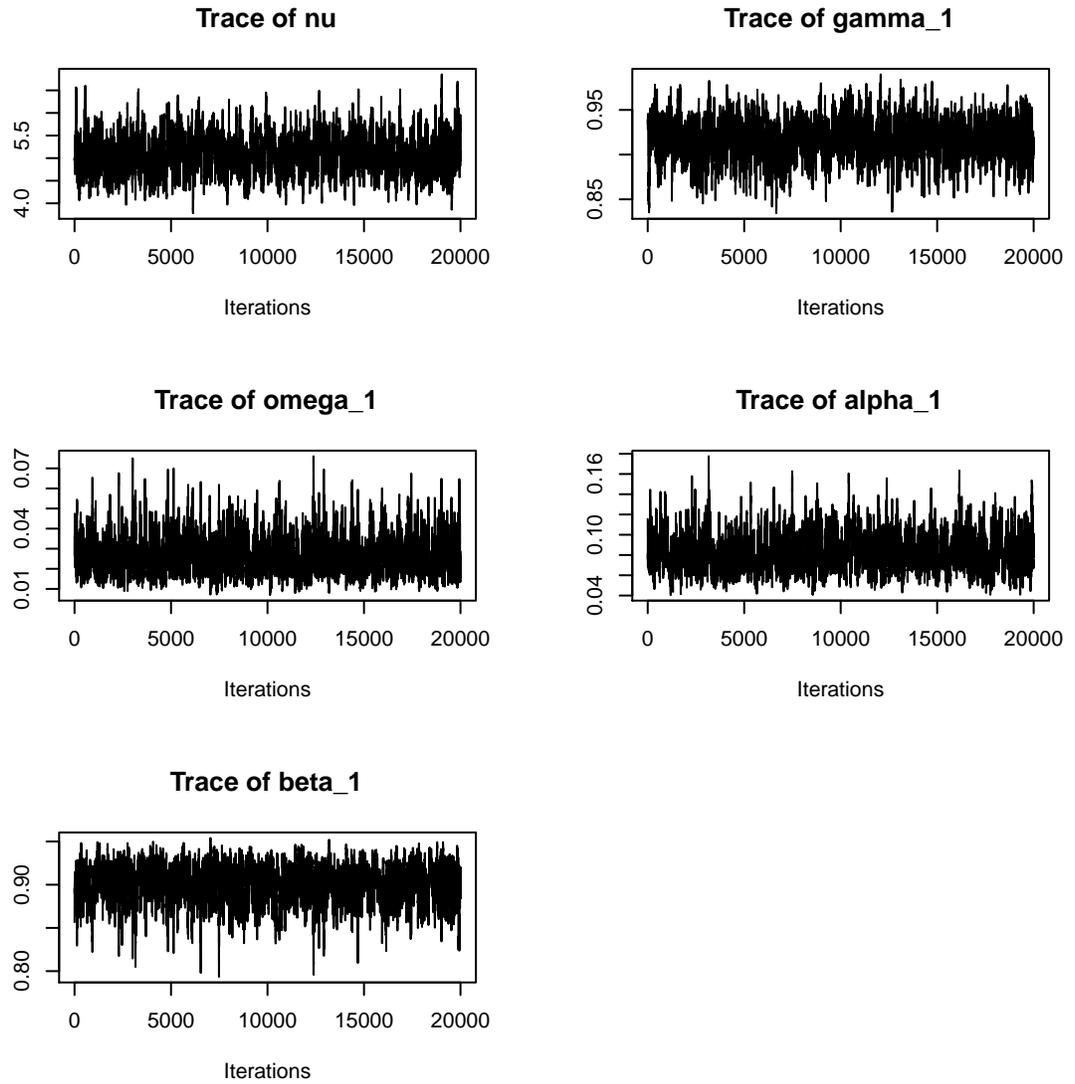}
\caption{Trace plot of simulated parameters in the GARCH(1,1) with
  Skew Student error distribution.}
\end{figure}

\begin{figure}
\includegraphics{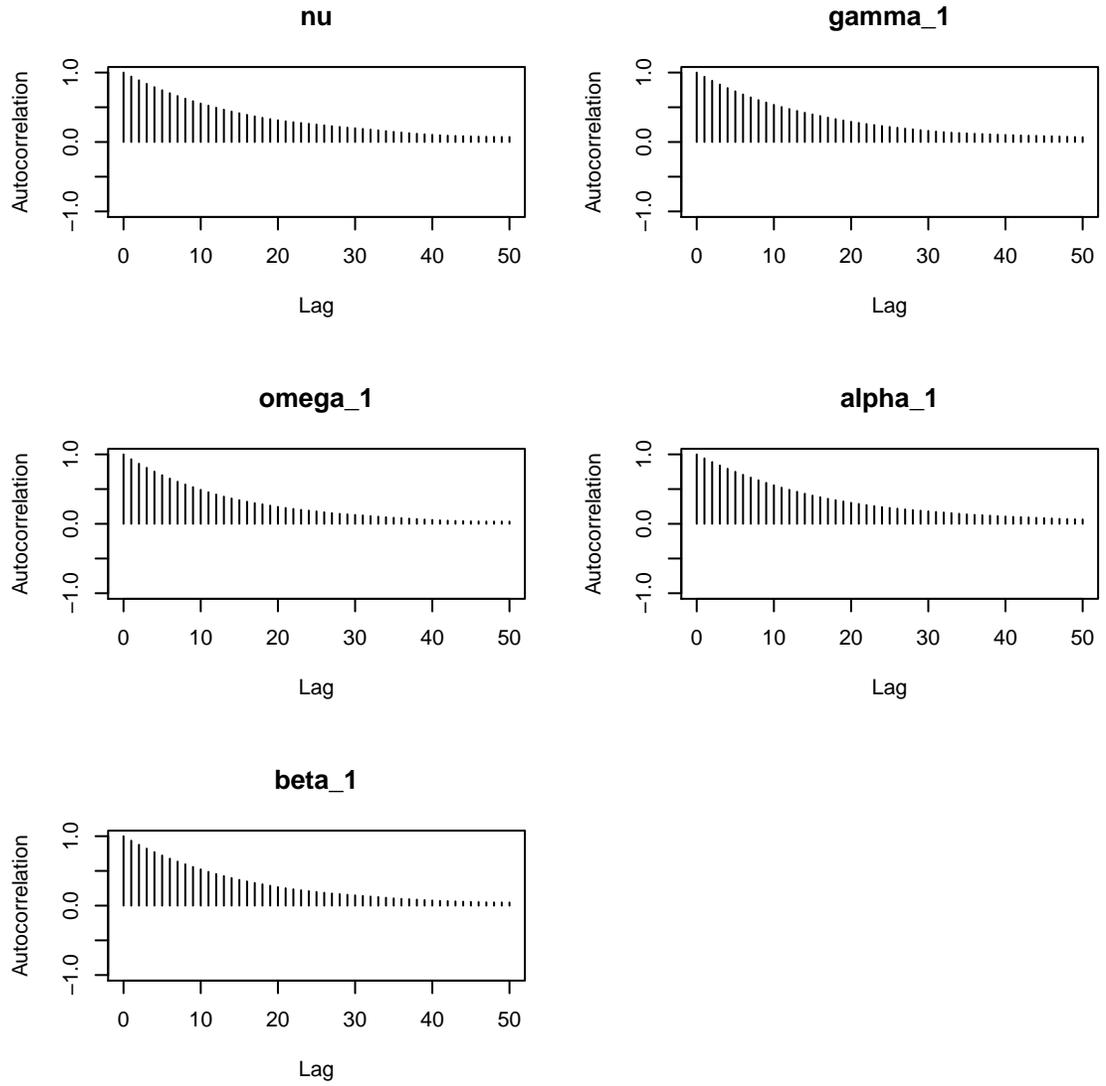}
\caption{Sample autocorrelations of simulated parameters in the GARCH(1,1) with
  Skew Student error distribution.}
\end{figure}

\begin{figure}
\includegraphics{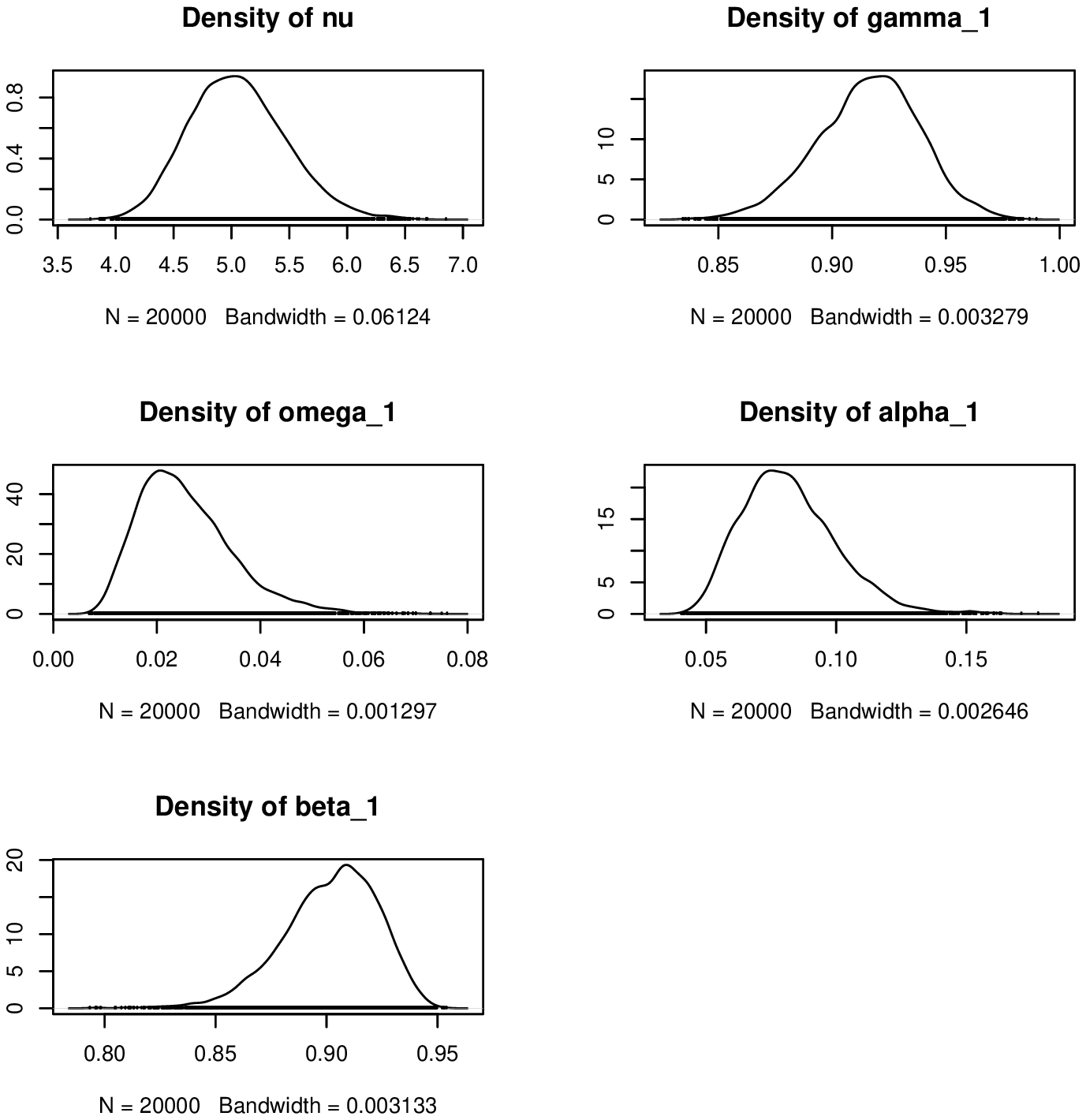}
\caption{Density estimates of parameters in the GARCH(1,1) with
  Skew Student error distribution.}
\end{figure}

\begin{figure}
\includegraphics{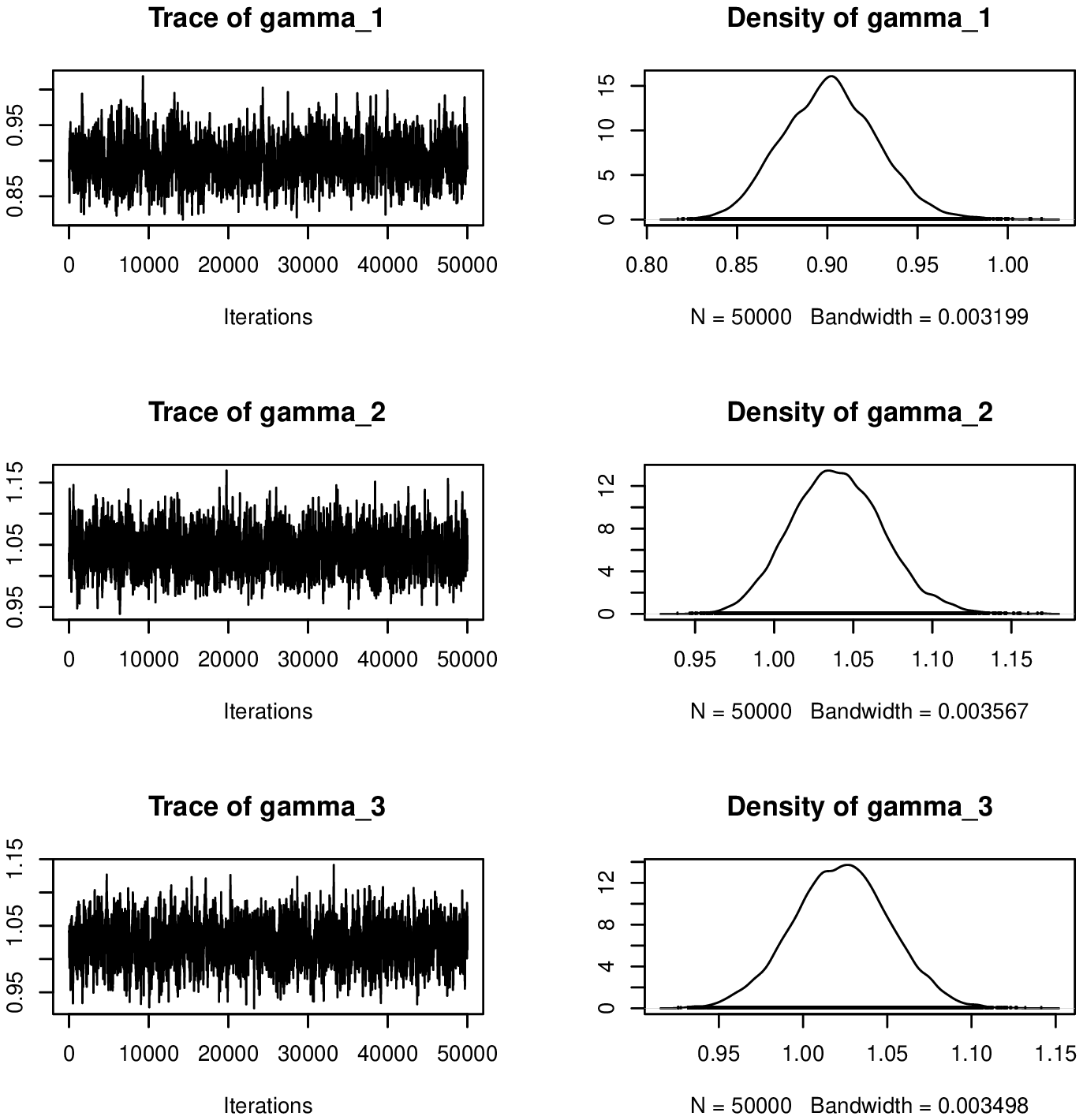}
\caption{Trace plot and posterior density estimates of skewness
  parameters for the DAX, CAC40 and NIKKEI indices using the skew
  mutivariate $t$ distribution.}
\label{fig:skewness}
\end{figure}

We now perform a multivariate analysis using the three daily indices,
DAX, CAC and NIKEEI. Figure \ref{fig:skewness} shows the estimated posterior densities of
the skewness parameters $\gamma_i$. They clearly indicate skewness in
the marginal distributions for the  DAX and CAC40 indices while symmetry
is more likely for the NIKKEI index.

\section{Conclusion}\label{sec:conclusion}

The current version of {\bf bayesDccGarch} uses {\tt R} and {\tt C}
and is available for the {\tt R} system for statistical computing
(\citeNP{R06}) from the Comprehensive R Archive Network at 
http://CRAN.R-project.org/. 

\section*{Acknowledgements}

The work of the first author was funded by CAPES - Brazil. The work of
the second author was supported by FAPESP - Brazil, under grant number
2011/22317-0. 


\end{document}